\documentclass[aps,pre,twocolumn,superscriptaddress,nofootinbib,longbibliography]{revtex4-2}

\usepackage{amsmath,amssymb}
\usepackage{booktabs}
\usepackage{array}
\usepackage{bm}
\usepackage{graphicx}
\usepackage[hidelinks]{hyperref}
\usepackage{microtype}

\usepackage{xcolor}

\begin{document}

\title{A memory-based three-state model of competing technology adoption: substitution regimes, multi-homing, and churn}

\author{Stefano Scialla}
\email{stefano.scialla@unicampus.it}
\affiliation{Department of Engineering, Universit\`a Campus Bio-Medico di Roma, 00128 Rome, Italy}
\author{Marco Patriarca}
\affiliation{Department of Computer Science, Aalto University School of Science, P.O. Box 11000, 00076 Aalto, Finland}
\author{Els Heinsalu}
\affiliation{School of Digital Technologies, Tallinn University, Tallinn, Estonia}
\author{Julyan H. E. Cartwright}
\affiliation{Instituto Andaluz de Ciencias de la Tierra, CSIC, 18100 Armilla, Spain}
\affiliation{Instituto Carlos I de F\'isica Te\'orica y Computacional, Universidad de Granada, 18071 Granada, Spain}

\date{\today}

\begin{abstract}
Technologies, products, platforms, and behavioral routines often compete through processes in which adoption is gradual, continued use requires reinforcement, and users may temporarily maintain more than one option. We formulate a homogeneous, well-mixed, three-state agent-based model of competition between an incumbent option \(X\) and a challenger option \(Y\). Agents are exclusive users of \(X\), exclusive users of \(Y\), or dual adopters \(Z\) who have both options available. Adoption is memory-based: an exclusive user adds the alternative only after accumulating enough adoption-relevant encounters within a finite learning window. Retention is also memory-based: a dual adopter continues to use both options only if each is sufficiently reinforced within a finite retention window. The central result is that this microscopic mechanism can reproduce aggregate usage signatures analogous to the four Adner--Kapoor technology-substitution regimes---creative destruction, robust coexistence, the illusion of resilience, and robust resilience---even though the model does not explicitly represent ecosystem emergence, ecosystem extension, complementors, prices, or strategic investment. Starting from the same small challenger seed, the four benchmark simulations differ only in post-entry mechanisms: adoption burden, retention burden, post-adoption usage preference, and the teaching role of dual adopters. Rolling usage shares reproduce aggregate substitution patterns analogous to the Adner--Kapoor regimes, while state-resolved trajectories and phase portraits reveal the underlying microscopic pathways. The results show that similar market-level substitution curves do not need to have unique causal interpretations. Ecosystem mechanisms may be essential in many empirical cases, but user-level finite-memory learning and retention alone can generate qualitatively similar aggregate regimes. The model therefore provides a compact baseline for linking technology-substitution trajectories to observable individual-level adoption, multi-homing, and discontinuance processes.
\end{abstract}

\maketitle

\section{Introduction}
\label{sec:intro}







Technological innovation is widely regarded as a principal driver of long-term economic and social change \cite{Bettencourt2007,Solow1957}. 
Yet its  impact is not fixed at the moment of invention: scientific insights and new ideas must first pass through research and development and then through a  process of the market, by which they reach and spread among users. Casimir describes this relationship as the \textit{science-technology spiral}, in which technological application follows scientific advance, often with a delay of years, while technology in turn reshapes the direction of science itself \cite{Casimir1983}. 
In the present paper, we study the latter stage of this process, i.e., what determines the pace and pattern of adoption of a technology that has moved beyond the laboratory and entered the market.

Technologies, platforms, and behavioral routines (``habits'') often compete for users' time, attention, and coordination opportunities. The societal impact of a technology is not determined solely by invention, but by adoption: who adopts, when they adopt, whether they keep using it, and whether a competing alternative survives \cite{Rogers2003}.

A recurring empirical feature is that adoption is typically not a one-shot switch. Many technologies require onboarding, practice, and repeated exposure before use becomes reliable and comfortable \cite{Rogers2003,JaspersonCarterZmud2005}. Conversely, abandonment (``churn'') often occurs after a sustained period of low engagement: skills decay, interfaces are forgotten, and routines fail to self-reinforce \cite{Bhattacherjee2001,ArthurBennettStanushMcNelly1998,WoodNeal2007,Lally2010}. These observations suggest that adoption dynamics have an explicit time-scale dimension: adoption requires repeated adoption-relevant experiences within a finite learning horizon and state changes depend on experiences accumulated over time, not only on instantaneous incentives. In addition, once a technology or habit has been adopted, it requires reinforcement to be maintained, i.e., technology adoption and retention are memory-based processes.
This connects to basic cognitive psychology and to the classical works of Ebbinghaus on learning \cite{Ebbinghaus-1885a,Murre2015}.

A broad literature models innovation diffusion at an aggregate level. A classic example is the Bass model, which combines external influence (e.g., advertising) and internal influence (word-of-mouth) to generate S-shaped adoption curves \cite{Bass1969}. Related network-based models distinguish advertising from word-of-mouth in innovation spreading \cite{KocsisKun2011}.
The structure of the Bass model has been influential and has been used beyond the diffusion of technological innovations, becoming a paradigmatic backbone for many models in complex systems and social physics, e.g., in opinion and language dynamics \cite{Castellano-2009a}, in conjunction with different microscopic models, starting with social reinterpretations of the basic Ising model \cite{Slanina,SznajdWeronSznajd2000}. 
Not coincidentally, the Bass model has also been used to study the diffusion of new words \cite{Ghanbarnejad2014}.

The diffusion-of-innovations tradition emphasizes adopter heterogeneity, social communication channels, and stages of adoption \cite{Rogers2003}. These approaches are valuable for fitting macroscopic adoption trajectories, but they typically compress individual learning and retention into a small number of rates.

At the micro level, two modeling traditions are especially relevant. The first is threshold models, where adoption occurs only after sufficient social reinforcement (e.g., a large enough fraction of neighbors are adopters) \cite{Granovetter1978}. The second is contagion (sometimes named ``complex contagion'') where repeated exposures may be needed for a behavior to spread \cite{CentolaMacy2007}.

However, many threshold/contagion models trigger adoption based on the current configuration of peers, rather than explicitly representing the possibility that learning requires multiple trials and that partial progress may be lost if not reinforced quickly enough. In other words, adoption is frequently treated as memoryless at the individual level. When time scales are important (onboarding, skill acquisition, habit formation), it would make sense to build memory into the microscopic rules.

Technology diffusion often involves competition between alternatives (standards, platforms, communication tools, work practices). Economic models of network effects and increasing returns highlight the possibility of tipping and lock-in: early advantages can become self-reinforcing \cite{KatzShapiro1985,Arthur1989}. Platform markets also feature multi-sided feedback and strategic interactions \cite{RochetTirole2003,Armstrong2006}. 
                                                  
For simplicity, in this study, we assume that there are two competing alternatives: an older technology, or incumbent, and a new technology, or challenger.

A particularly important phenomenon is multi-homing: users adopt and maintain multiple platforms, tools, or routines simultaneously \cite{LandsmanStremersch2011}. Multi-homing can play two conceptually different roles:
\begin{itemize}
\item[(i)] \textit{Bridge/compatibility effect}: if multi-homers can interact with users of either option, they may mitigate coordination failures and support interactions across otherwise separated user groups without requiring everyone to adopt both technologies \cite{DoganogluWright2006}.
\item[(ii)] \textit{Teaching/exposure effect}: because multi-homers have direct experience with the alternative, they may demonstrate it, share relevant know-how, or otherwise increase others' likelihood of adoption \cite{FangWuClough2021,AralWalker2011}.
\end{itemize}
Both mechanisms have precedents in the literature, although they are generally studied separately, and their relative importance is likely to depend on the context. In this paper we build a deliberately minimal baseline that includes both a bridge role and a teaching role: dual adopters accommodate exclusive users, but may also expose them to the alternative.

In habit research, repeated practice in stable contexts is often emphasized as a driver of automaticity, while discontinuation weakens routines \cite{WoodNeal2007,Lally2010}. In technology adoption, onboarding and switching costs similarly imply that adoption requires sustained effort. These observations motivate a model in which:
\begin{itemize}
\item[(i)] Adoption is cumulative: an exclusive user adopts the alternative only after enough adoption-relevant events (exposures/trials) accumulate.
\item[(ii)] Cumulative progress can decay: if progress is not reinforced within a finite horizon, the attempt can be abandoned (reset).
\item[(iii)] Retention requires reinforcement: dual users remain dual only if both options are used sufficiently often within a retention window; otherwise the less-used option is dropped.
\end{itemize}
This finite-window approach is phenomenological, but it yields a clear separation between how fast exposure happens, controlled by contact patterns and interaction rates, and how demanding adoption and retention are, controlled by thresholds and time windows.

The above described features of the technology adoption process show that it presents strong analogies with the competition between two languages. Language competition models typically consider three linguistic states: monolingual speaker of language $X$, monolingual speaker of language $Y$, and bilingual state $Z$. In the technology competition framework, these states can be reinterpreted, respectively, as exclusive use of option $X$ (incumbent), exclusive use of option $Y$ (challenger), and dual use $Z$ (multi-homing). As previously explained, our aim is to model technology competition as a memory-based framework where adoption and abandonment occur over finite time windows as a consequence of sufficient, or insufficient, repeated exposures and reinforcements. To achieve this goal, we adapt a three-state, memory-based language-competition framework developed in previous work \cite{scialla2023}, in which language learning and attrition are not instantaneous transitions but finite-window processes: a monolingual speaker becomes bilingual only after accumulating a sufficient number of interactions in the other language within a learning interval, while a bilingual speaker retains both languages only if both are used sufficiently often within a maintenance interval.

Having defined this microscopic analogy, we need a macroscopic reference against which to interpret the simulated adoption trajectories. Technology substitution can proceed in qualitatively different ways: a challenger may rapidly displace an incumbent, coexist with it for a long period, appear blocked for a long time before suddenly taking over, or fail to displace it within the relevant observation horizon. To compare the model simulations with such typical real-life substitution scenarios, we use the framework developed by Adner and Kapoor \cite{adner2016smj,adner2016hbr} as an external benchmark.

Adner and Kapoor study the pace of technological substitution by emphasizing that competition between an old and a new technology does not depend only on the intrinsic performance of the two alternatives. It also depends on the broader innovation ecosystem surrounding them. In their terminology, the new technology may face an \emph{emergence challenge}, meaning that complementary components, infrastructure, or user capabilities needed for its effective use may still be underdeveloped. At the same time, the old technology may have an \emph{extension opportunity}, meaning that improvements in its own ecosystem can prolong its viability. Different combinations of these two forces generate four characteristic substitution regimes.

The first regime, \emph{creative destruction}, corresponds to rapid replacement: the challenger faces limited obstacles to emergence, while the incumbent has little room for further extension. Note that here and throughout, ``creative destruction'' refers to the first Adner--Kapoor regime we just described, not to the broader Schumpeterian concept of creative destruction as a mechanism of sustained macroeconomic growth \cite{Schumpeter1942,AghionHowitt1992}. The second Adner--Kapoor regime, \emph{robust coexistence}, occurs when the challenger can make early progress but the incumbent also remains viable, producing an extended period in which both technologies retain material market presence. The third, the \emph{illusion of resilience}, describes cases in which the incumbent appears stable for a long time because the challenger initially faces substantial barriers, but once those barriers are overcome, substitution becomes rapid. The fourth, \emph{robust resilience}, corresponds to the slowest substitution path: the challenger faces persistent obstacles and the incumbent continues to benefit from strong extension opportunities, so that dominance by the challenger is absent or strongly delayed.

Illustrative cases discussed by Adner and Kapoor include the rapid replacement of dot-matrix by inkjet printers (creative destruction), the prolonged coexistence of hybrid and conventional internal-combustion engines (robust coexistence), the delayed but then rapid displacement of paper maps by GPS navigation (illusion of resilience), and the prolonged resistance of barcodes to replacement by RFID (radio-frequency identification), a case of robust resilience \cite{adner2016hbr}.

In the present paper, we do not attempt to model the full ecosystem mechanisms identified by Adner and Kapoor. Our model does not include complementors, strategic investment, prices, infrastructure, or endogenous technological improvement. Instead, we use the four regimes as aggregate trajectory benchmarks. The question is whether a much smaller set of user-level mechanisms---finite-memory adoption, post-adoption retention, multi-homing, and repeated reinforcement---can generate usage trajectories that resemble these well-known substitution patterns. This distinction is important, as similar aggregate curves can arise from different microscopic histories. For example, a long coexistence period may reflect persistent multi-homing, separate groups of exclusive users, or repeated cycles of trial and abandonment. Likewise, an incumbent plateau may indicate that the challenger is not being tried at all, or that it is being tried but not retained. The model therefore records not only aggregate usage shares, but also state-resolved quantities such as dual adoption, reversion, completed switching, and failed onboarding attempts. These observables allow us to ask whether similar market-level trajectories are produced by distinct individual-level pathways.

\section{Model}
\label{sec:model}

\subsection{Population, states, and initial condition}

We consider a population of $N$ agents. Each agent is always in exactly one of three states:
\begin{itemize}
\item $X$: exclusive user of the incumbent option $X$;
\item $Y$: exclusive user of the challenger option $Y$;
\item $Z$: dual adopter or multi-homer who uses both options.
\end{itemize}
Let $n_\ell(t)$ be the number of agents in state $\ell\in\{X,Y,Z\}$ at discrete time $t$. The corresponding fractions are
\begin{equation}
N_\ell(t)=\frac{n_\ell(t)}{N}, \qquad N_X(t)+N_Y(t)+N_Z(t)=1.
\end{equation}
The incumbent--challenger experiment starts without dual adopters and with a small challenger seed:
\begin{equation}
N_X(0)=1-y_0,\qquad N_Y(0)=y_0,\qquad N_Z(0)=0,
\label{eq:seed}
\end{equation}
where $0<y_0<1/2$. In the benchmark simulations below, the same $y_0$ is used in all panels. The seed represents a small initial population of real challenger users, not a large established challenger market. In empirical terms, such a seed may be interpreted as the early adopter base generated by launch advertising, promotional offers, pilot programs, institutional introduction, media visibility, or other initial market-entry activities that place the challenger into actual use before subsequent diffusion proceeds through user interactions.

The dynamics evolve with a discrete time, which is discrete and may be interpreted as days, weeks, or another application-specific interval. 
We assume a simple all-to-all topology for the underlying network, as here we want to focus on how memory affects the adoption process. Thus, at each time step, two agents are picked up randomly and let interact with each other.
We denote by \(N_{\rm int}\) the interaction intensity, defined as the average number of pairwise interaction events in which an agent participates during one time step. Thus, at each time step the simulation samples
\begin{equation}
N_{\rm pairs}=\frac{N_{\rm int}N}{2}
\end{equation}
pairs of distinct agents uniformly from the population. The factor \(1/2\) appears because each pairwise event involves two agents. The value \(N_{\rm int}\) is therefore not a fixed degree or a fixed number of contacts assigned to every individual in every time step; it is a population-level average interaction rate, and the same agent may be sampled more than once during a time step. Here and in what follows, an interaction, encounter or meeting is interpreted broadly as a communication episode, shared workflow, coordination task, observation, trial, or other encounter in which use of one option can generate experience and reinforcement. Realized use is generated only by these pairwise interactions.

\subsection{Microscopic variables and memory counters}

Each agent $i$ stores a primary option $L_i^{(1)}\in\{X,Y\}$. A dual adopter also stores a secondary option $L_i^{(2)}\in\{X,Y\}$ with $L_i^{(2)}\neq L_i^{(1)}$; for exclusive users $L_i^{(2)}=0$. The primary option is the agent's default or more established routine and is retained as a tie-breaking convention when neither option is sufficiently reinforced.

An exclusive adopter carries two learning variables: a counter $k_i(t)$ of adoption-relevant events accumulated toward adding the alternative, and a learning clock $\tau_i^K(t)$ measuring elapsed time since the current learning attempt began. The learning clock starts with the first adoption-relevant event and remains zero while $k_i=0$.

A dual adopter carries two option-specific usage counters, $m_{X,i}(t)$ and $m_{Y,i}(t)$, and one retention clock $\tau_i^M(t)$. The option-specific counters are used because post-adoption retention can be asymmetric even when the retention window is common.

\subsection{Interaction rules}

The encounter rules determine both the option used in each sampled pair and the corresponding updates of learning, retention, and usage counters. They are described below and summarized in Table~\ref{tab:encounter_rules}.

\paragraph{Exclusive--exclusive encounters.}
If two exclusive adopters use the same option, $X+X$ or $Y+Y$, the common option is used and no learning counter changes. If an $X$-exclusive and a $Y$-exclusive agent meet, the encounter is adoption-relevant for both:
\begin{equation}
k_i\rightarrow k_i+1,\qquad k_j\rightarrow k_j+1.
\end{equation}

\paragraph{Exclusive--dual encounters.}
If an exclusive adopter meets a dual adopter, the interaction uses the exclusive adopter's option; the dual adopter accommodates. The corresponding usage counter of the dual adopter is incremented. For example, in an $X+Z$ encounter,
\begin{equation}
m_{X,j}\rightarrow m_{X,j}+1
\end{equation}
for the dual agent $j$. The exclusive user also receives an adoption-relevant event toward the alternative with probability $p_{\rm teach}$:
\begin{equation}
k_i\rightarrow k_i+1 \quad \hbox{with probability }p_{\rm teach}.
\end{equation}
Thus $p_{\rm teach}=0$ implements a pure bridge or compatibility effect, whereas $p_{\rm teach}=1$ means that every exclusive--dual encounter also teaches the alternative.

\paragraph{Dual--dual encounters.}
When two dual adopters meet, both options are available to both agents. The pair uses $Y$ with probability $q_Y(t)$ and $X$ with probability $1-q_Y(t)$:
\begin{equation}
P(\hbox{use }Y\mid Z+Z)=q_Y(t),\qquad P(\hbox{use }X\mid Z+Z)=1-q_Y(t).
\end{equation}
The selected option's usage counter is incremented for both agents. The neutral case is $q_Y=1/2$. Values $q_Y>1/2$ give the challenger a post-adoption usage advantage when both products are available, whereas $q_Y<1/2$ favors the incumbent. This parameter is intentionally distinct from adoption thresholds: a product may be easy to add but rarely selected after adoption, or difficult to add but strongly preferred once available.

\begin{table*}[t]
\caption{Encounter rules and usage bookkeeping in the well-mixed memory-based adoption model. Each sampled pair contributes two realized-use incidences. Learning counters apply only to exclusive adopters; retention counters apply only to dual adopters.}
\label{tab:encounter_rules}
\centering
\footnotesize
\setlength{\tabcolsep}{4pt}
\renewcommand{\arraystretch}{1.15}
\begin{tabular}{lll}
\toprule
\parbox[t]{0.16\textwidth}{\textbf{Encounter}} &
\parbox[t]{0.28\textwidth}{\textbf{Option used}} &
\parbox[t]{0.48\textwidth}{\textbf{Memory and usage updates}} \\
\midrule
\parbox[t]{0.16\textwidth}{$X+X$} &
\parbox[t]{0.28\textwidth}{$X$} &
\parbox[t]{0.48\textwidth}{Record two $X$ use incidences. No learning counter changes.} \\
\addlinespace
\parbox[t]{0.16\textwidth}{$Y+Y$} &
\parbox[t]{0.28\textwidth}{$Y$} &
\parbox[t]{0.48\textwidth}{Record two $Y$ use incidences. No learning counter changes.} \\
\addlinespace
\parbox[t]{0.16\textwidth}{$X+Y$} &
\parbox[t]{0.28\textwidth}{Each exclusive user uses or observes the other option} &
\parbox[t]{0.48\textwidth}{Record one $X$ and one $Y$ use incidence. Both agents receive one adoption-relevant event toward the alternative: $k_i\to k_i+1$, $k_j\to k_j+1$.} \\
\addlinespace
\parbox[t]{0.16\textwidth}{$X+Z$ or $Y+Z$} &
\parbox[t]{0.28\textwidth}{The exclusive user's option; the dual adopter accommodates} &
\parbox[t]{0.48\textwidth}{Record two use incidences for the exclusive user's option. Increment the dual adopter's corresponding retention counter, $m_X$ or $m_Y$. With probability $p_{\rm teach}$, increment the exclusive user's learning counter toward the alternative.} \\
\addlinespace
\parbox[t]{0.16\textwidth}{$Z+Z$} &
\parbox[t]{0.28\textwidth}{$Y$ with probability $q_Y(t)$ and $X$ with probability $1-q_Y(t)$} &
\parbox[t]{0.48\textwidth}{Record two use incidences for the selected option. Increment the selected option's retention counter for both dual adopters.} \\
\bottomrule
\end{tabular}
\end{table*}

\subsection{Usage bookkeeping and market-level observables}

Option use is recorded separately from repertoire state. The usage-bookkeeping convention for each encounter type is included in Table~\ref{tab:encounter_rules}. In an $X+X$ encounter both incidences are assigned to $X$; in a $Y+Y$ encounter both are assigned to $Y$; and in an $X+Y$ encounter one incidence is assigned to each option. In an exclusive--dual encounter, both incidences are assigned to the exclusive user's option because the dual adopter accommodates. In a dual--dual encounter, both incidences are assigned to the option selected by the $q_Y$ rule.

Let $I_X(t)$ and $I_Y(t)$ denote the daily numbers of use incidences. With a fixed number of sampled pairs,
\begin{equation}
I_X(t)+I_Y(t)=2N_{\rm pairs}.
\end{equation}
The state fractions $N_X,N_Y,N_Z$ describe repertoires; the usage incidences $I_X,I_Y$ describe realized use. This distinction is essential when dual adopters are common.

\subsection{Adoption within a finite learning horizon}

Let $T_K$ be the learning or onboarding horizon. We use a destination-based convention:
\begin{itemize}
\item $K_X$ is the number of adoption-relevant events required for a $Y$-exclusive agent to add $X$;
\item $K_Y$ is the number required for an $X$-exclusive agent to add $Y$.
\end{itemize}
A $Y$-exclusive agent becomes dual if
\begin{equation}
k_i\geq K_X \quad \hbox{and}\quad \tau_i^K\leq T_K,
\end{equation}
while an $X$-exclusive agent becomes dual if
\begin{equation}
k_i\geq K_Y \quad \hbox{and}\quad \tau_i^K\leq T_K.
\end{equation}
If the applicable threshold is not reached before the learning window expires, the attempt is abandoned:
\begin{equation}
k_i\rightarrow 0,\qquad \tau_i^K\rightarrow 0.
\end{equation}
When adoption succeeds, the original option remains primary, the added option becomes secondary, and the retention variables are initialized:
\begin{equation}
m_{X,i},m_{Y,i},\tau_i^M\rightarrow 0.
\end{equation}
The reciprocal quantities $a_X=1/K_X$ and $a_Y=1/K_Y$ may be interpreted as effective adoptabilities. Lower $K_\ell$ means that option $\ell$ is added after fewer relevant experiences, conditional on exposure. The learning-window cases, including successful adoption and failed onboarding, are summarized in Table~\ref{tab:transition_checks}.

\subsection{Retention, abandonment, and completed substitution}

Dual adoption is not automatically permanent. Let $T_M$ be the retention window, and let $M_X$ and $M_Y$ be the minimum numbers of uses needed to retain $X$ and $Y$. At the end of a retention window, four outcomes are possible: the agent remains dual if both retention conditions are met; becomes $X$-exclusive if only the $X$ condition is met; becomes $Y$-exclusive if only the $Y$ condition is met; or retains the primary option and drops the secondary if neither condition is met. These retention-window outcomes, together with the corresponding reset rules and pathway interpretations, are summarized in Table~\ref{tab:transition_checks}.

Whenever an agent returns to exclusivity, the learning variables are reset. An $X\rightarrow Z\rightarrow Y$ trajectory represents completed substitution, whereas $X\rightarrow Z\rightarrow X$ represents trial followed by reversion.

The average reinforcement burdens are
\begin{equation}
r_X=\frac{M_X}{T_M},\qquad r_Y=\frac{M_Y}{T_M}.
\end{equation}
A smaller $r_\ell$ means that option $\ell$ is easier to sustain and is therefore more ``sticky.''

\begin{table*}[t]
\caption{End-of-window transition checks. The learning check applies to exclusive adopters, while the retention check applies to dual adopters. The table also identifies the individual-level pathways that underlie aggregate substitution curves.}
\label{tab:transition_checks}
\centering
\footnotesize
\setlength{\tabcolsep}{4pt}
\renewcommand{\arraystretch}{1.15}
\begin{tabular}{lll}
\toprule
\parbox[t]{0.17\textwidth}{\textbf{Agent state and check}} &
\parbox[t]{0.31\textwidth}{\textbf{Condition}} &
\parbox[t]{0.45\textwidth}{\textbf{Outcome and reset}} \\
\midrule
\parbox[t]{0.17\textwidth}{Exclusive user, active learning attempt} &
\parbox[t]{0.31\textwidth}{$k_i<K_{\rm dest}$ and $\tau_i^K<T_K$} &
\parbox[t]{0.45\textwidth}{The learning attempt continues. The learning clock advances while $k_i>0$.} \\
\addlinespace
\parbox[t]{0.17\textwidth}{Exclusive user, successful learning} &
\parbox[t]{0.31\textwidth}{$k_i\ge K_{\rm dest}$ and $\tau_i^K\le T_K$, where $K_{\rm dest}=K_Y$ for an $X$-exclusive agent and $K_X$ for a $Y$-exclusive agent} &
\parbox[t]{0.45\textwidth}{The agent becomes dual. The original option remains primary, the added option becomes secondary, and $m_{X,i},m_{Y,i},\tau_i^M$ are initialized to zero.} \\
\addlinespace
\parbox[t]{0.17\textwidth}{Exclusive user, failed onboarding} &
\parbox[t]{0.31\textwidth}{$k_i<K_{\rm dest}$ when the learning window expires} &
\parbox[t]{0.45\textwidth}{The attempt is abandoned and $k_i,\tau_i^K\to0$. This records failed onboarding without dual adoption.} \\
\addlinespace
\parbox[t]{0.17\textwidth}{Dual adopter, both options retained} &
\parbox[t]{0.31\textwidth}{$m_{X,i}\ge M_X$ and $m_{Y,i}\ge M_Y$ at $\tau_i^M\ge T_M$} &
\parbox[t]{0.45\textwidth}{The agent remains dual and the retention counters are reset for the next window.} \\
\addlinespace
\parbox[t]{0.17\textwidth}{Dual adopter, only $X$ retained} &
\parbox[t]{0.31\textwidth}{$m_{X,i}\ge M_X$ and $m_{Y,i}<M_Y$} &
\parbox[t]{0.45\textwidth}{The agent becomes $X$-exclusive. For an incumbent-origin user this is $X\to Z\to X$ reversion; for a challenger-origin user it is completed reverse switching.} \\
\addlinespace
\parbox[t]{0.17\textwidth}{Dual adopter, only $Y$ retained} &
\parbox[t]{0.31\textwidth}{$m_{X,i}<M_X$ and $m_{Y,i}\ge M_Y$} &
\parbox[t]{0.45\textwidth}{The agent becomes $Y$-exclusive. For an incumbent-origin user this is completed substitution, $X\to Z\to Y$.} \\
\addlinespace
\parbox[t]{0.17\textwidth}{Dual adopter, neither option retained} &
\parbox[t]{0.31\textwidth}{$m_{X,i}<M_X$ and $m_{Y,i}<M_Y$} &
\parbox[t]{0.45\textwidth}{The agent retains the primary option and drops the secondary option. Learning variables are reset when the agent returns to exclusivity.} \\
\bottomrule
\end{tabular}
\end{table*}

\subsection{Homogeneity}

For simplicity and in order to strictly compare the microscopic model rules to the external benchmark, the model contains no fixed agent heterogeneity. In particular, there are no agent-specific values $K_{X,i}$, $K_{Y,i}$, or $T_{M,i}$, and no dispersion parameters $\sigma_K$ or $\sigma_M$. All agents share the same thresholds at a given time. The only differences between agents arise from stochastic interaction histories and from the state changes produced by those histories.

\section{Heuristic exposure and retention rates}

For an $X$-exclusive agent, a meeting with a $Y$-exclusive agent always contributes one event toward adding $Y$, while a meeting with a dual adopter contributes with probability $p_{\rm teach}$. The expected rate of adoption-relevant events is therefore
\begin{equation}
\lambda_{X\rightarrow Z}(t)\simeq N_{\rm int}\left[N_Y(t)+p_{\rm teach}N_Z(t)\right].
\label{eq:lambda}
\end{equation}
The corresponding accumulation-time estimate is
\begin{equation}
t_{X\rightarrow Z}\simeq \frac{K_Y}{N_{\rm int}\left(N_Y+p_{\rm teach}N_Z\right)}.
\end{equation}
At the start of the benchmark experiments, $N_Z(0)=0$ and $N_Y(0)=y_0$, so the first-order learning condition is
\begin{equation}
y_0 \gtrsim \theta_Y, \qquad \theta_Y\equiv \frac{K_Y}{N_{\rm int}T_K}.
\label{eq:seedcondition}
\end{equation}
If $\theta_Y\ll y_0$, challenger adoption begins quickly. If $\theta_Y$ is close to or slightly above $y_0$, most learning attempts fail, but rare successful attempts in a finite population may create a small dual population. If $p_{\rm teach}>0$ and $Y$ is retained once adopted, this small dual population can become the seed of a later cascade. If $\theta_Y$ is far above $y_0$ or if post-adoption retention of $Y$ is weak, the incumbent may remain dominant throughout the observation horizon.

Retention can be interpreted similarly. If a dual adopter's pairwise interactions use $Y$ with probability $\psi_Y(t)$, approximate reinforcement rates are
\begin{align}
\omega_X(t)&\simeq N_X(t)+N_Z(t)\left[1-\psi_Y(t)\right],\\
\omega_Y(t)&\simeq N_Y(t)+N_Z(t)\psi_Y(t),
\end{align}
where $\psi_Y(t)=q_Y(t)$ for dual--dual encounters under the product-choice rule, with additional $Y$ use coming from meetings with $Y$-exclusive agents. Retention requires roughly
\begin{equation}
N_{\rm int}T_M\omega_X\gtrsim M_X,\qquad N_{\rm int}T_M\omega_Y\gtrsim M_Y.
\end{equation}
Thus $K_\ell$ controls conversion of exposure into adoption, while $M_\ell/T_M$ and $q_Y$ control post-adoption survival and realized use.



\section{Results and discussion}
\label{sec:results}

The model is evaluated by pattern-oriented comparison rather than by matching a single endpoint \cite{grimm2005}. The comparison has two levels. At the aggregate level, we ask whether the model can reproduce four qualitative substitution signatures derived from the Adner--Kapoor framework: creative destruction, robust coexistence, the illusion of resilience, and robust resilience. At the state-resolved level, we ask which microscopic pathways generate those aggregate signatures. In particular, we distinguish completed switching, reversion, persistent multi-homing, and failed onboarding, which are not visible in a two-share usage curve.

All benchmark simulations use the same population size, interaction rate, memory windows, observation horizon, and initial challenger seed:
\[
N=1000,\qquad N_{\rm int}=16,\qquad T_K=730,\qquad T_M=120,
\]
\[
t_{\rm fin}=7000,\qquad y_0=0.05,\qquad K_X=50{,}000.
\]
The large value of \(K_X\) makes reverse adoption of \(X\) by \(Y\)-exclusive users negligible, so the benchmark atlas focuses on incumbent-to-challenger dynamics. The common seed gives the challenger the same small initial realized usage share in every panel. Thus, differences among the panels arise from post-entry mechanisms---adoption burden, retention burden, post-adoption usage preference, and teaching by dual adopters---rather than from different launch times or different initial challenger shares.

In the simulations reported below, the finite-memory conditions are evaluated using moving temporal windows. At each time step, the learning counter \(k_i(t)\) counts adoption-relevant events that occurred during the most recent \(T_K\) time steps; events older than \(T_K\) no longer contribute to the adoption condition. Similarly, for dual adopters, \(m_{X,i}(t)\) and \(m_{Y,i}(t)\) count \(X\)- and \(Y\)-use events during the most recent \(T_M\) time steps, and retention is evaluated using these moving-window counts. This implementation is used for the microscopic learning and retention rules. The rationale for this choice is that adoption and retention are driven by recent exposure and recent use: events that occurred far in the past should progressively lose relevance for current learning or maintenance. The moving-window formulation therefore represents finite memory as the gradual expiration of old events, avoiding an artificial periodic reset of counters while preserving \(T_K\) and \(T_M\) as the characteristic learning and retention horizons.

All plotted trajectories are ensemble medians over independent stochastic realizations. For Figs.~\ref{fig:atlas} and~\ref{fig:states}, each benchmark is simulated over 50 independent replicates with the common parameters listed above. For the phase portraits in Fig.~\ref{fig:phase}, each initial condition is simulated over 30 independent replicates, using \(N=300\) agents for computational efficiency. In all cases, medians are taken pointwise in time.

Table~\ref{tab:benchmarks} summarizes the intended qualitative mapping between aggregate substitution signatures and reduced-form model mechanisms. The conditions are hypotheses about parameter regions, not deterministic classifications and not literal ecosystem mechanisms.

\begin{table*}[t]
\caption{Four literature-grounded technology-substitution benchmarks and corresponding reduced-form model mechanisms in the homogeneous initial-seed model. The same \(y_0\), \(N\), \(N_{\rm int}\), \(T_K\), \(T_M\), and \(t_{\rm fin}\) are used across the usage and state-trajectory panels.}
\label{tab:benchmarks}
\centering
\footnotesize
\setlength{\tabcolsep}{4pt}
\renewcommand{\arraystretch}{1.15}
\begin{tabular}{llll}
\toprule
\parbox[t]{0.15\textwidth}{\textbf{Benchmark}} &
\parbox[t]{0.25\textwidth}{\textbf{Aggregate usage signature}} &
\parbox[t]{0.28\textwidth}{\textbf{Reduced-form model mechanism}} &
\parbox[t]{0.23\textwidth}{\textbf{State-resolved diagnostic}} \\
\midrule
\parbox[t]{0.15\textwidth}{B1. Creative destruction} &
\parbox[t]{0.25\textwidth}{Early challenger takeoff, rapid usage dominance, and a narrow transition from incumbent to challenger use.} &
\parbox[t]{0.28\textwidth}{\(\theta_Y\ll y_0\), strong \(Y\) retention, \(q_Y>1/2\), and an incumbent retention burden high enough that dual adopters eventually drop \(X\).} &
\parbox[t]{0.23\textwidth}{Short transient \(Z\) wave and high probability of completed substitution, \(X\rightarrow Z\rightarrow Y\).} \\
\addlinespace
\parbox[t]{0.15\textwidth}{B2. Robust coexistence} &
\parbox[t]{0.25\textwidth}{The challenger makes early inroads, while both options retain material usage for an extended interval; full displacement does not occur within the observation horizon.} &
\parbox[t]{0.28\textwidth}{\(\theta_Y\ll y_0\), durable dual adoption, and post-adoption usage preference close enough to neutrality that both options remain materially used.} &
\parbox[t]{0.23\textwidth}{Persistent \(Z\) population, long dual lifetimes, and sustained coexistence in realized use.} \\
\addlinespace
\parbox[t]{0.15\textwidth}{B3. Illusion of resilience} &
\parbox[t]{0.25\textwidth}{A long incumbent plateau is followed by steep challenger takeoff and rapid usage-share inversion.} &
\parbox[t]{0.28\textwidth}{\(\theta_Y\) close to the seed boundary, strong \(Y\) retention or usage advantage, and positive teaching by dual adopters.} &
\parbox[t]{0.23\textwidth}{Failed or rare learning events before takeoff, followed by a surge in \(Z\) and completed switches.} \\
\addlinespace
\parbox[t]{0.15\textwidth}{B4. Robust resilience} &
\parbox[t]{0.25\textwidth}{The incumbent remains dominant for the whole horizon, or challenger dominance is right-censored.} &
\parbox[t]{0.28\textwidth}{\(\theta_Y\) near or above the seed boundary, strong incumbent retention, weak challenger retention or \(q_Y<1/2\), and low teaching.} &
\parbox[t]{0.23\textwidth}{Trial may occur, but \(X\rightarrow Z\rightarrow X\) reversion dominates and durable challenger dominance is not reached.} \\
\bottomrule
\end{tabular}
\end{table*}

The specific parameter values corresponding to each technology-substitution benchmark according to Adner-Kapoor (from B1 to B4) are reported in Table~\ref{tab:parameters}. These values specify numerical choices inserted into the microscopic encounter and transition rules summarized in Tables~\ref{tab:encounter_rules} and~\ref{tab:transition_checks}; they are not additional model rules. All parameters reported in Table~\ref{tab:parameters} are time independent.

\begin{table*}[t]
\caption{Illustrative benchmark parameters used for Figs.~\ref{fig:atlas}--\ref{fig:reduced_mosaic}. Figs.~\ref{fig:atlas} and~\ref{fig:states} use the common seed \(y_0=0.05\); Fig.~\ref{fig:phase} uses the same benchmark rules while varying the initial challenger share along the edge \(N_X(0)+N_Y(0)=1\), \(N_Z(0)=0\); Fig.~\ref{fig:reduced_mosaic} represents the same parameters in reduced coordinates. All parameters reported here are time independent.}
\label{tab:parameters}
\centering
\footnotesize
\setlength{\tabcolsep}{4pt}
\renewcommand{\arraystretch}{1.15}
\begin{tabular}{lcccccl}
\toprule
\textbf{Benchmark} &
\(K_Y\) &
\(M_X\) &
\(M_Y\) &
\(q_Y\) &
\(p_{\rm teach}\) &
\parbox[t]{0.32\textwidth}{\textbf{Mechanistic role}} \\
\midrule
B1 &
260 &
500 &
60 &
0.85 &
0.00 &
\parbox[t]{0.32\textwidth}{Fast adoption and durable displacement. The challenger is easy to add, easy to retain, and strongly favored after adoption.} \\
\addlinespace
B2 &
260 &
500 &
500 &
0.47 &
0.00 &
\parbox[t]{0.32\textwidth}{Persistent robust coexistence. The challenger is easy to add, both options are easy to retain, and post-adoption usage preference remains close to neutrality.} \\
\addlinespace
B3 &
650 &
500 &
50 &
0.85 &
0.30 &
\parbox[t]{0.32\textwidth}{Near-boundary learning delay followed by a teaching-supported cascade. Adoption from the seed is initially difficult, but dual adopters can teach and the challenger is easy to retain once adopted.} \\
\addlinespace
B4 &
620 &
300 &
900 &
0.30 &
0.02 &
\parbox[t]{0.32\textwidth}{Limited trial and right-censored incumbent dominance. The challenger is near the learning boundary, weakly taught, disfavored in dual--dual use, and difficult to retain.} \\
\bottomrule
\end{tabular}
\end{table*}

The three figures provide complementary views of the same benchmark construction. Figure~\ref{fig:atlas} shows realized use, Fig.~\ref{fig:states} shows the state composition behind realized use, and Fig.~\ref{fig:phase} shows state-space trajectories from a range of initial challenger shares. We discuss the four benchmarks separately within each of these three representations.

\subsection{Aggregate usage signatures}
\label{subsec:fig1_usage}

The benchmark comparison is based first on realized option use. To reduce stochastic fluctuations in the daily interaction counts, we plot rolling usage shares over a window of \(h\) time steps:
\begin{equation}
\bar{s}^{(h)}_Y(t)=
\frac{\sum_{\tau=\max(0,t-h+1)}^{t} I_Y(\tau)}
{\sum_{\tau=\max(0,t-h+1)}^{t}\left[I_X(\tau)+I_Y(\tau)\right]},
\label{eq:rolling_usage}
\end{equation}
with
\begin{equation}
\bar{s}^{(h)}_X(t)=1-\bar{s}^{(h)}_Y(t).
\end{equation}
We use \(h=21\) time steps in all simulations. If one time step is interpreted as one day, this corresponds to a three-week rolling window. This window is much shorter than the selected learning horizon, \(T_K=730\), and is used only for visualization; it does not enter the microscopic transition rules.

In all time-series plots, time is reported in dimensionless form as \(t/T_K\), so that one unit on the horizontal axis corresponds to one full learning or onboarding horizon. Since we select \(T_K=730\), \(t/T_K=1\) corresponds to 730 time steps, or approximately two years if one time step is interpreted as one day. This normalization is used only for visualization and comparison across panels; the microscopic simulation itself is performed in discrete time steps.

\begin{figure*}[t]
\centering
\includegraphics[width=0.98\textwidth]{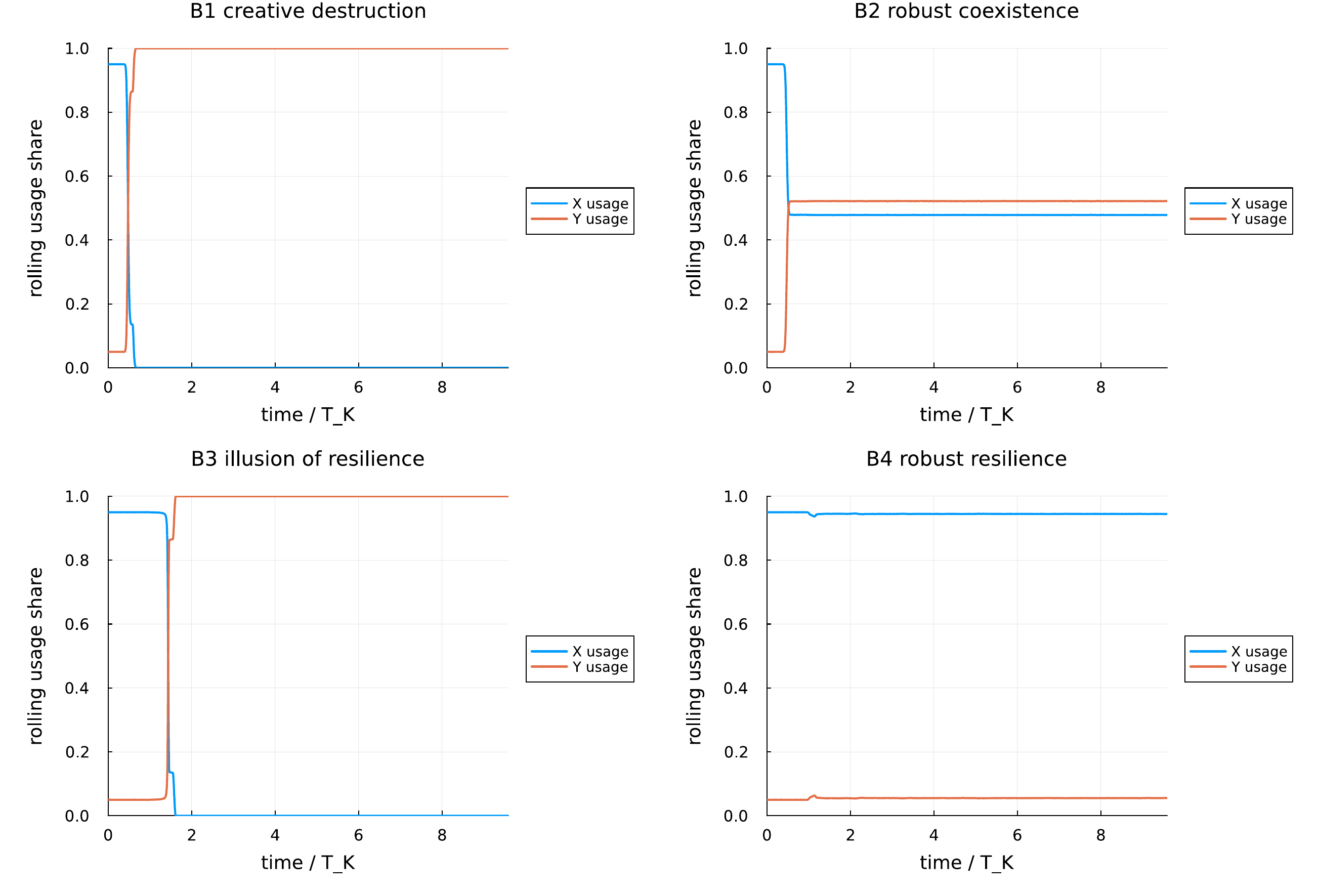}
\caption{Four benchmark signatures with a common initial challenger seed. Curves show ensemble-median rolling usage shares for the incumbent \(X\) and challenger \(Y\) as functions of time measured in units of the learning window \(T_K\). All panels use the same initial seed \(N_Y(0)=0.05\), no initial dual adopters, homogeneous thresholds, pairwise realized use, and moving-window finite memory. Parameter differences are listed in Table~\ref{tab:parameters}.}
\label{fig:atlas}
\end{figure*}

Figure~\ref{fig:atlas} shows that the same initial challenger seed can lead to qualitatively different aggregate usage trajectories. In B1, the challenger rapidly displaces the incumbent. The transition occurs early and over a narrow interval, after which \(Y\) accounts for almost all realized use. This is the aggregate signature of creative destruction.

B2 differs from B1 because the challenger becomes materially used early, but the system then remains in a broad coexistence band. The usage shares settle close to parity, with \(Y\) slightly above \(X\), and neither option eliminates the other over the observation horizon. This is the aggregate signature of robust coexistence in the present benchmark: the central feature is sustained material use of both technologies, not eventual complete replacement.

B3 ends in challenger dominance, like B1, but its aggregate trajectory is different from B1. The incumbent remains close to complete usage dominance for more than one learning-window time scale before a rapid inversion occurs. Because the challenger seed is present from the beginning, the plateau does not mean that \(Y\) is absent from the system. Rather, the aggregate curve indicates that early exposure is insufficient to generate a macroscopic cascade until the reinforcing feedback through dual adopters becomes strong enough.

B4 is the only panel in which the challenger does not achieve aggregate dominance. The incumbent remains dominant throughout the observation horizon, while \(Y\) remains close to the seed share, with only small deviations associated with limited trial and dual adoption. This is the aggregate signature of robust resilience. At the level of usage shares alone, however, it is not possible to tell whether the challenger is absent from the adoption process or whether it is tried but not sufficiently reinforced. That distinction requires the state-resolved representation in Fig.~\ref{fig:states}.

\subsection{State-resolved trajectories}
\label{subsec:fig2_states}

\begin{figure*}[t]
\centering
\includegraphics[width=0.98\textwidth]{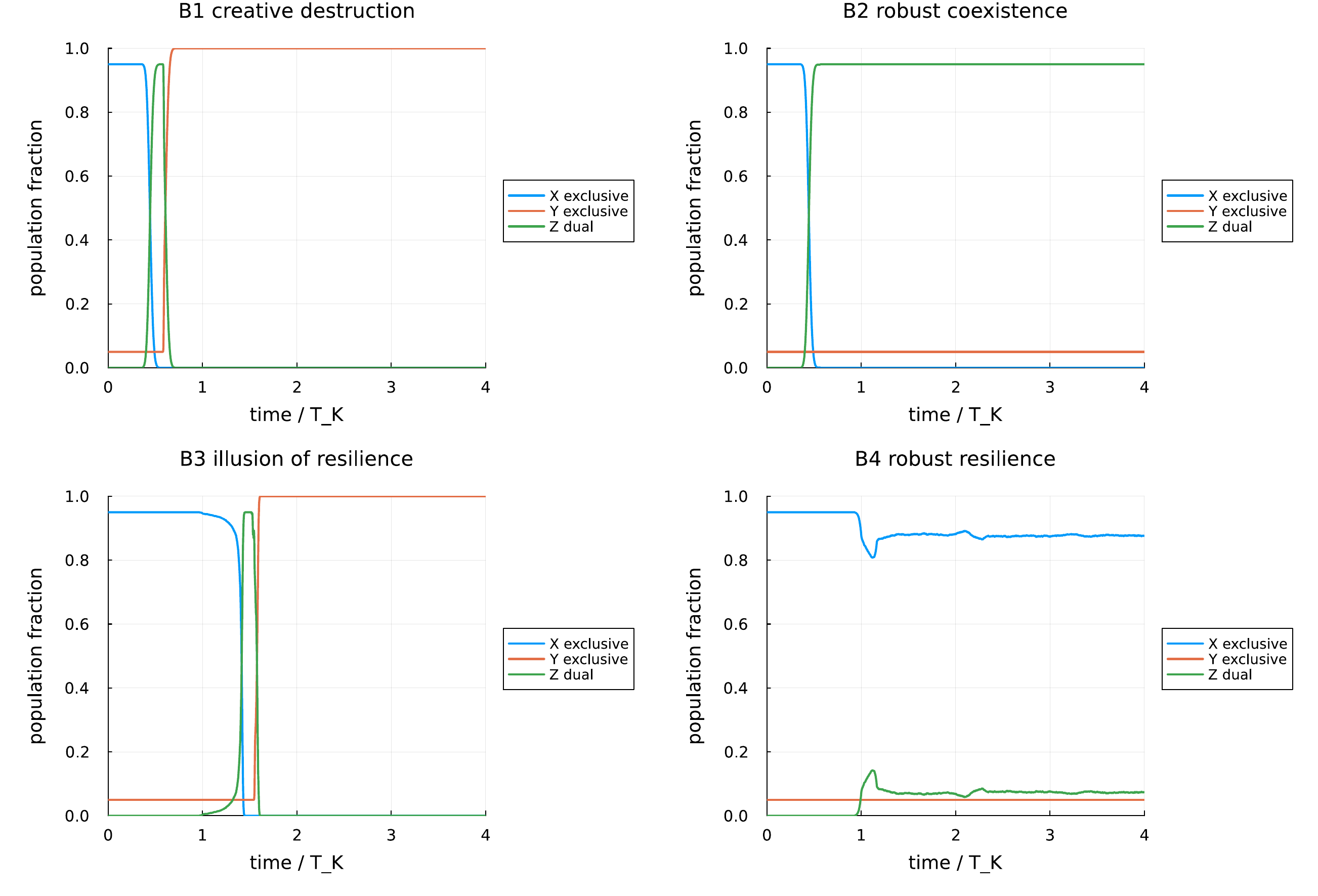}
\caption{State-resolved trajectories underlying the four benchmark signatures. Curves show ensemble-median population fractions of exclusive incumbent users \(N_X(t)\), exclusive challenger users \(N_Y(t)\), and dual adopters \(N_Z(t)\) for the same simulations shown in Fig.~\ref{fig:atlas}. The horizontal axis is restricted to the first four learning windows to make the state transitions visible. The figure distinguishes whether aggregate usage patterns are generated by completed switching, persistent multi-homing, or limited trial without durable challenger dominance under the moving-window finite-memory implementation.}
\label{fig:states}
\end{figure*}

The rolling shares in Fig.~\ref{fig:atlas} measure realized use. They are distinct from the state fractions \(N_X(t)\), \(N_Y(t)\), and \(N_Z(t)\), which describe the composition of user repertoires. Figure~\ref{fig:states} reports the corresponding ensemble-median state fractions for the same benchmark simulations. The horizontal axis is restricted to the range \(0\leq t/T_K\leq4\) in order to resolve the formation and disappearance of dual-adoption waves.

The state trajectories show the mechanisms behind aggregate usage curves. In B1, the transition from incumbent to challenger passes through a large but short-lived dual-adopter wave. The population first moves from \(X\)-exclusive use into \(Z\), and then from \(Z\) into \(Y\)-exclusive use. Thus, the B1 usage curve is generated by completed substitution: dual adoption is a transient bridge rather than a persistent state.

In B2, the state-level behavior is very different. The early movement away from \(X\)-exclusive use produces a large and persistent dual-adopter population. The initial \(Y\)-exclusive seed remains present, but most former incumbent-exclusive users enter \(Z\) and remain there. Coexistence in realized use is therefore sustained primarily by multi-homing rather than by two large exclusive subpopulations. This is the key distinction between B2 and B1 at the state level: both involve early adoption of \(Y\), but in B1 the dual state is short-lived, whereas in B2 the dual state persists and supports material use of both options.

In B3, the state trajectories reveal the hidden mechanism behind the long incumbent plateau in Fig.~\ref{fig:atlas}. For much of the early period, most agents remain \(X\)-exclusive, with only a small dual population. When the teaching feedback becomes large enough, \(N_Z(t)\) rises sharply and then collapses as agents complete the transition to \(Y\)-exclusive use. The aggregate inversion is therefore produced by a delayed but intense dual-adoption wave.

In B4, the state trajectories show that robust resilience does not necessarily mean absence of trial. The \(Y\)-exclusive seed remains present, and a small dual-adopter population appears, but this dual adoption is not amplified into a cascade. Most agents remain \(X\)-exclusive, and the challenger does not become the main realized-use option. The state trajectory therefore indicates limited trial and low-level multi-homing without durable challenger dominance.

The contrast between B3 and B4 is especially informative. Both place the challenger near the learning boundary, but the fate of early dual adopters differs. In B3, teaching and strong post-adoption reinforcement amplify rare successes into a cascade. In B4, weak teaching, low challenger preference, and difficult challenger retention prevent the same feedback from taking hold. The aggregate distinction between illusion of resilience and robust resilience is therefore produced by the fate of early dual adopters, not by the timing of challenger entry.

\subsection{State-space phase portraits}
\label{subsec:fig3_phase}

\begin{figure*}[t]
\centering
\includegraphics[width=0.98\textwidth]{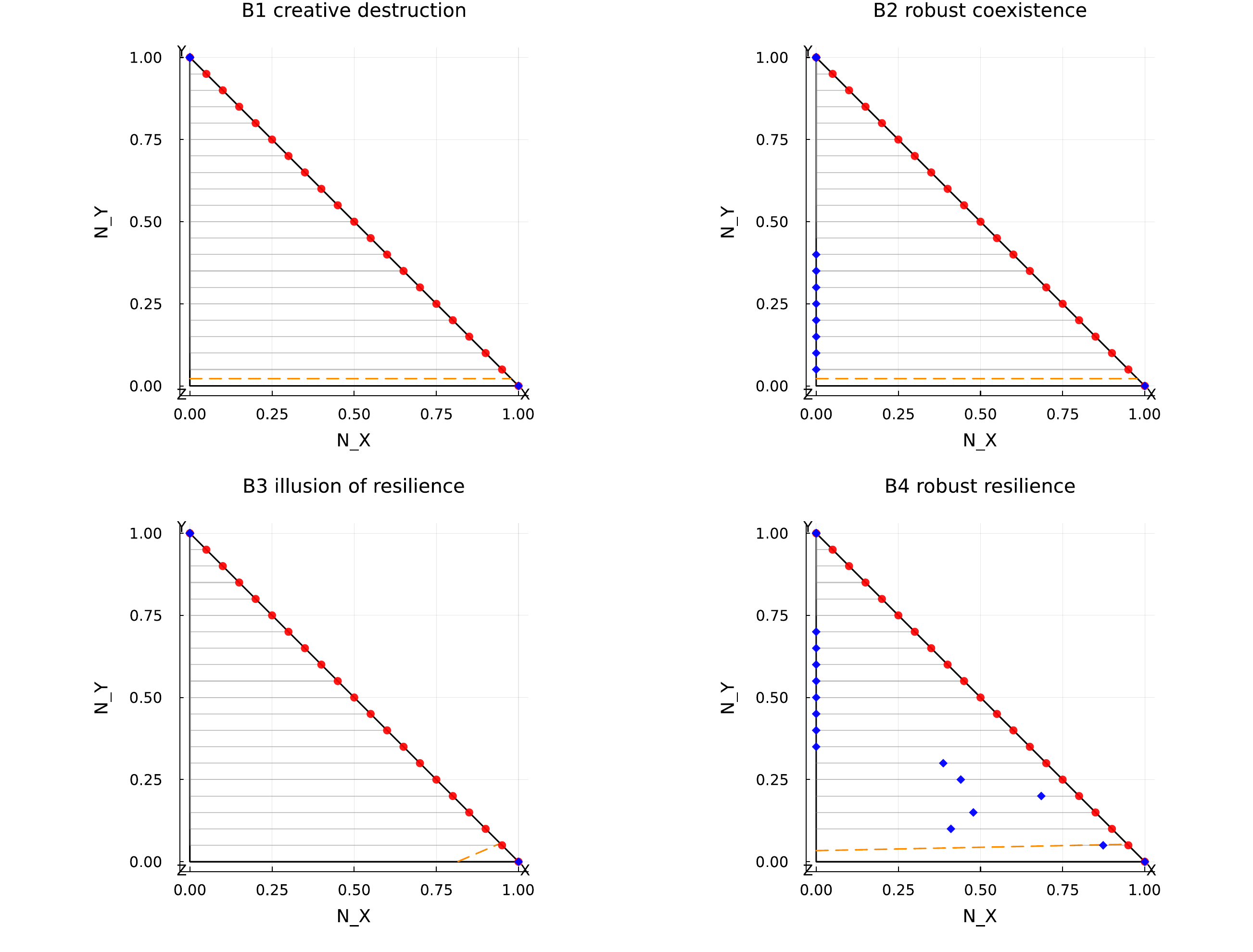}
\caption{State-space phase portraits corresponding to the four benchmark regimes. The triangular domain is the simplex \(N_X+N_Y+N_Z=1\). Red points denote initial conditions with \(N_Z(0)=0\), blue points denote final ensemble-median states, and gray curves show the corresponding trajectories in the \((N_X,N_Y)\) plane. The diagonal \(N_X+N_Y=1\) corresponds to the absence of dual adopters, while points below the diagonal have \(N_Z>0\). The dashed line marks the approximate early-learning boundary \(N_Y+p_{\rm teach}N_Z=\theta_Y\).}
\label{fig:phase}
\end{figure*}

Figure~\ref{fig:phase} gives a third view of the same mechanisms. Instead of fixing the initial challenger share at \(y_0=0.05\), the phase portraits simulate each benchmark rule from a range of initial challenger shares along the edge \(N_X(0)+N_Y(0)=1\), \(N_Z(0)=0\). Trajectories are plotted in the \((N_X,N_Y)\) plane. The diagonal \(N_X+N_Y=1\) corresponds to \(N_Z=0\), while points below the diagonal have a positive dual-adopter fraction. Red points denote initial conditions, blue points denote final states, and gray curves show the state-space trajectories. The dashed line marks the approximate early-learning boundary
\[
N_Y+p_{\rm teach}N_Z=\theta_Y,
\qquad
\theta_Y=\frac{K_Y}{N_{\rm int}T_K},
\]
which indicates where \(X\)-exclusive agents can accumulate enough exposure to add the challenger within one learning window.

The B1 phase portrait shows a large basin of attraction toward the challenger corner. Once the initial challenger share is large enough to cross the learning boundary, trajectories rapidly leave the initial edge and end near \(Y\)-exclusive dominance. The intermediate dual state is traversed quickly, consistent with the short \(Z\) wave in Fig.~\ref{fig:states}.

The B2 phase portrait is qualitatively different. Many trajectories end away from both exclusive corners and close to the \(N_X=0\) edge, where the population contains the original \(Y\)-exclusive seed together with a large dual-adopter component. This confirms that the coexistence signature is associated with persistent multi-homing: most incumbent users add the challenger, but the system does not collapse into exclusive challenger use. The phase portrait therefore makes visible a mechanism that the usage panel alone cannot identify.

The B3 phase portrait highlights the threshold-like character of the illusion-of-resilience case. Initial conditions below the effective learning boundary remain near the incumbent-dominated edge, while initial conditions that generate enough dual adoption move toward challenger dominance. This sharp separation reflects the positive feedback between dual adoption and teaching: once the \(Z\) population is large enough, it increases the exposure of remaining \(X\)-exclusive users and produces a cascade.

The B4 phase portrait shows the opposite outcome. Many trajectories remain incumbent dominated or move only partially into the interior of the simplex. Some initial conditions generate trial and temporary or low-level dual adoption, but the final states do not concentrate near the challenger corner. Thus, robust resilience is not simply a lack of exposure; it can also be a regime in which exposure and trial occur but are not reinforced strongly enough to produce durable challenger dominance.

\begin{figure*}[t]
\centering
\includegraphics[width=0.98\textwidth]{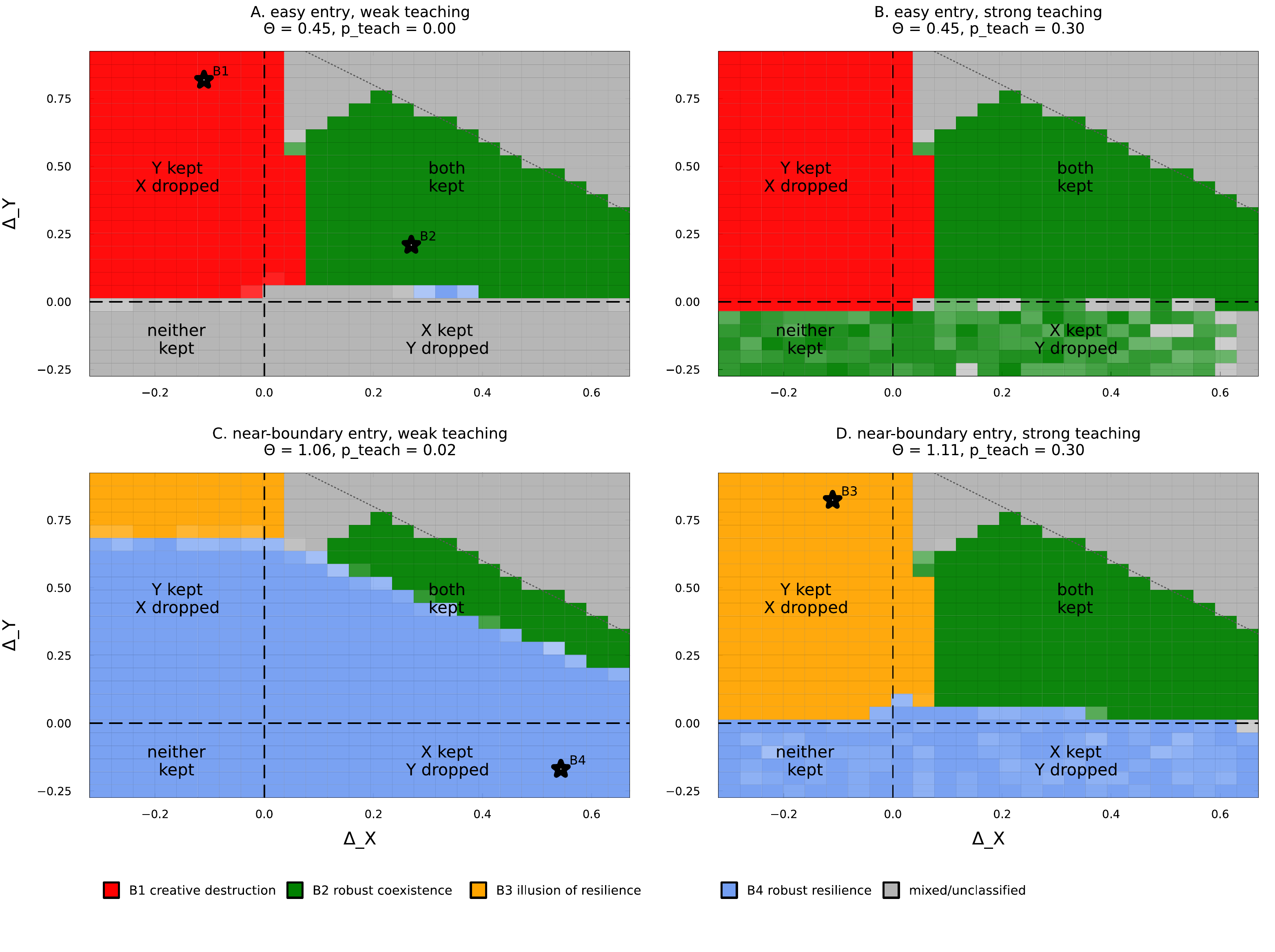}
\caption{Reduced-coordinate regime mosaic. Each panel shows a simulation-based regime map in the \((\Delta_X,\Delta_Y)\) plane for fixed seed-normalized entry burden \(\Theta\) and teaching probability \(p_{\rm teach}\). The reduced coordinates are
\(\Theta=K_Y/(N_{\rm int}T_Ky_0)\),
\(\Delta_X=(1-q_Y)-M_X/(N_{\rm int}T_M)\), and
\(\Delta_Y=q_Y-M_Y/(N_{\rm int}T_M)\).
The quadrant labels give the schematic retention interpretation in the dual-dominated limit: \(\Delta_X>0\) means that \(X\) can be retained, while \(\Delta_Y>0\) means that \(Y\) can be retained. Colors denote the modal simulated regime classification: B1 creative destruction, B2 robust coexistence, B3 illusion of resilience, B4 robust resilience, or mixed/unclassified. Color intensity indicates the fraction of stochastic realizations assigned to the modal regime. Stars mark the benchmark parameter sets reported in Table~\ref{tab:parameters}. The vertical and horizontal dashed lines indicate \(\Delta_X=0\) and \(\Delta_Y=0\), respectively. The diagonal dotted line marks the feasibility boundary \(\Delta_X+\Delta_Y=1\); points above this line do not correspond to non-negative retention burdens for any admissible value of \(q_Y\).}
\label{fig:reduced_mosaic}
\end{figure*}

\subsection{Correspondence between benchmark regimes and parameter space regions}
\label{subsec:parameterspace}

Beyond the specific parameter combinations summarized in Table \ref{tab:parameters} and analyzed in detail in previous sections, it is interesting to study the correspondence between the four Adner-Kapoor benchmark regimes and the parameter space regions in our model. To do this, we define a set of reduced coordinates using the heuristic exposure and retention estimates. The first of these coordinates is the challenger adoption burden relative to the initial seed,
\begin{equation}
\Theta=\frac{\theta_Y}{y_0}
=\frac{K_Y}{N_{\rm int}T_K y_0}.
\label{eq:Theta}
\end{equation}
Values \(\Theta<1\) indicate that the initial challenger seed is large enough, at mean-field level, to provide the exposure needed for \(X\)-exclusive agents to add \(Y\) within one learning window. Values \(\Theta\simeq 1\) place the system near the seed-entry boundary, where stochastic successful adoption events and teaching by dual adopters become important.

The other two reduced variables describe post-adoption retention. Let us introduce the retention burdens
\begin{equation}
\rho_X=\frac{M_X}{N_{\rm int}T_M},
\qquad
\rho_Y=\frac{M_Y}{N_{\rm int}T_M}.
\label{eq:rho_retention}
\end{equation}
In the dual-dominated limit, \(N_Z\rightarrow 1\), the approximate reinforcement probabilities are \(\omega_X\simeq 1-q_Y\) and \(\omega_Y\simeq q_Y\). This suggests the retention margins
\begin{equation}
\Delta_X=(1-q_Y)-\rho_X,
\qquad
\Delta_Y=q_Y-\rho_Y.
\label{eq:retention_margins}
\end{equation}
A positive \(\Delta_X\) means that the incumbent can be retained by dual adopters in this limit, whereas a positive \(\Delta_Y\) means that the challenger can be retained. The four quadrants of the \((\Delta_X,\Delta_Y)\) plane therefore have a direct interpretation: \(Y\) kept and \(X\) dropped for \(\Delta_X<0,\Delta_Y>0\); both options kept for \(\Delta_X>0,\Delta_Y>0\); \(X\) kept and \(Y\) dropped for \(\Delta_X>0,\Delta_Y<0\); and neither option comfortably retained for \(\Delta_X<0,\Delta_Y<0\). The diagonal line \(\Delta_X+\Delta_Y=1\) is a feasibility boundary of the reduced coordinates.

Figure~\ref{fig:reduced_mosaic} uses these variables to summarize the simulated regimes. Each panel is a simulation-based map in the \((\Delta_X,\Delta_Y)\) plane, conditioned on a value of \(\Theta\) and \(p_{\rm teach}\). Colors denote the modal simulated regime classification over independent stochastic realizations, while color intensity reflects the fraction of realizations assigned to that modal regime. The classification is based on the aggregate usage and state diagnostics used throughout the paper: early final challenger dominance is classified as B1, delayed final challenger dominance as B3, persistent coexistence with a large dual-adopter population as B2, and continued incumbent dominance with low challenger usage as B4. Points that do not satisfy these criteria are classified as mixed or unclassified.

The top row, where \(\Theta=0.45\), corresponds to easy entry from the initial seed. In this case, teaching is not essential: comparing the weak-teaching and strong-teaching panels shows that the main organization is provided by the retention margins. The quadrant \(\Delta_X<0,\Delta_Y>0\), where the challenger is retained but the incumbent is not, is dominated by B1-like creative destruction. The quadrant \(\Delta_X>0,\Delta_Y>0\), where both options can be retained, is dominated by B2-like coexistence. Thus, once entry is easy, the distinction between displacement and coexistence is controlled primarily by whether dual adopters retain one option or both.

The lower-left panel combines near-boundary entry with weak teaching. Here the map is dominated by B4-like robust resilience over a broad region. This occurs because \(\Theta\simeq 1\) makes the initial seed insufficient to generate widespread adoption, and weak teaching prevents the small dual-adopter population from amplifying exposure. In this regime, the retention margins may be favorable to the challenger in principle, but they become relevant only if enough agents actually enter the dual state.

The lower-right panel shows that stronger teaching can reopen the displacement pathway near the seed-entry boundary. When \(p_{\rm teach}=0.30\), a B3-like region appears in the \(\Delta_X<0,\Delta_Y>0\) quadrant. This is the reduced-coordinate signature of the illusion-of-resilience mechanism: the challenger is initially close to the learning boundary, but once a sufficient dual-adopter population appears, teaching raises the effective exposure of remaining incumbent-exclusive users and triggers a delayed cascade.

The Table~\ref{tab:parameters} benchmark points are consistent with this organization. B1 and B3 lie in the retention quadrant where \(Y\) is kept and \(X\) is dropped, but they are separated by the entry burden and the presence of teaching. B2 lies in the both-kept quadrant, consistent with persistent multi-homing. B4 lies in the incumbent-kept, challenger-dropped quadrant under near-boundary entry and weak teaching. The reduced-coordinate mosaic therefore supports the interpretation that the four benchmark signatures are not arbitrary parameter choices: they are organized by seed-normalized entry burden, retention margins, and teaching by dual adopters. At the same time, the reduction is approximate, because \(\Delta_X\) and \(\Delta_Y\) are derived in the dual-dominated limit; deviations from the quadrant schematic reflect the finite transition dynamics and the fact that \(N_Z\) is not always close to one.

\section{Conclusion}
\label{sec:conclusion}

The central result of this study is that the four Adner--Kapoor substitution regimes can be reproduced, at the level of aggregate usage trajectories, by a microscopic model based on finite-memory adoption and retention. In the Adner--Kapoor framework, creative destruction, robust coexistence, the illusion of resilience, and robust resilience arise from different combinations of ecosystem emergence challenges for the new technology and extension opportunities for the old technology. The present model does not explicitly represent those ecosystem mechanisms. It contains no complementors, infrastructure layers, strategic investment, price dynamics, or endogenous product improvement. Nevertheless, the simulations show that analogous usage signatures can emerge from user-level mechanisms alone: repeated exposure, finite learning windows, post-adoption reinforcement, multi-homing, and abandonment after insufficient use.

This result does not imply that ecosystem explanations are unnecessary or incorrect. Rather, it shows that aggregate substitution curves are not mechanically unique. A curve that resembles creative destruction, coexistence, delayed takeoff, or resilience does not need to identify the underlying cause by itself. The same aggregate pattern may arise from ecosystem bottlenecks, but it may also arise from the microscopic timing of learning, trial, reinforcement, and discontinuance. The model therefore provides a complementary interpretation of the Adner--Kapoor regimes: instead of asking only whether a technology's supporting ecosystem is ready, it asks whether individual users receive enough repeated exposure to adopt, whether the adopted option is reinforced often enough to be retained, and whether dual adopters function mainly as bridges that protect the incumbent or as teachers that accelerate diffusion.

The four benchmark simulations illustrate this point. In the creative-destruction case, the challenger is easy to add, easy to retain, and strongly favored after adoption. The state-resolved trajectory shows that dual adoption is only a short-lived intermediate state: users move from incumbent exclusivity to dual use and then quickly to challenger exclusivity. In the robust-coexistence case, by contrast, the challenger also makes early inroads, but the dual state persists. Coexistence is therefore not simply a flat aggregate curve; it is sustained by a large multi-homing population that continues to generate realized use of both options. In the illusion-of-resilience case, the challenger is present from the beginning but remains near the learning boundary. The incumbent appears stable for a long time because most users do not yet accumulate sufficient adoption-relevant exposure. Once a small dual-adopter population generates enough teaching feedback, however, the system undergoes a rapid cascade. Finally, in the robust-resilience case, trial may occur but is not reinforced strongly enough to become durable adoption. The typical microscopic pathway is reversion from dual use back to incumbent exclusivity.

The state-resolved trajectories and phase portraits are essential for this interpretation. The rolling usage shares show whether the aggregate path resembles one of the four benchmark regimes, but they do not reveal how that path is produced. The same apparent coexistence curve could reflect persistent multi-homing, stable exclusive subpopulations, or repeated trial and abandonment. Likewise, an incumbent plateau could mean that the challenger is never tried, or that it is repeatedly tried but not retained. By tracking \(N_X(t)\), \(N_Y(t)\), \(N_Z(t)\), and trajectories in the \(X\)--\(Y\)--\(Z\) simplex, the model separates usage signatures from the individual pathways that generate them. This is the main added value of the framework: it turns aggregate substitution regimes into hypotheses about observable microscopic processes.

The phase portraits also clarify the role of initial conditions and learning boundaries. Because adoption requires accumulated exposure within a finite window, the initial challenger seed matters through its relation to the effective threshold
$\theta_Y={K_Y} / {N_{\rm int}T_K}$.
When the seed and subsequent exposure are sufficient relative to this threshold, incumbent users can enter the dual state and further diffusion becomes possible. When the system remains near or below the boundary, adoption may be delayed, blocked, or dependent on teaching by the few dual adopters that appear. Thus, even in a homogeneous well-mixed population, the outcome depends not only on the final attractiveness of the challenger but also on the timing and accumulation of exposure.

The reduced-coordinate regime mosaic reinforces this interpretation by showing that the benchmark signatures are not isolated or finely tuned parameter points: in the scanned \((\Theta,\Delta_X,\Delta_Y,p_{\rm teach})\) space, they occupy extended and largely contiguous domains organized by the seed-entry condition and by the signs of the retention margins.

The model should therefore be interpreted as a post-entry, interaction-driven baseline. The small initial challenger seed represents the outcome of launch activity, early advertising, institutional introduction, promotional offers, pilot programs, or early adopters already present in the market. After this initial condition, the baseline deliberately asks how far user-level mechanisms alone can carry the diffusion process. In this sense, an interaction should be understood broadly: it may represent direct word of mouth, social-media exposure, online reviews, observation of use, technical help from peers, shared workflow requirements, referral messages, compatibility pressure, or any encounter in which another user's adoption generates information, trial, reinforcement, or coordination value.

This modeling choice is motivated by a substantial diffusion literature in which internal influence, word of mouth, peer effects, and social reinforcement are central mechanisms of adoption. The Bass model, for example, separates external influence, such as advertising, from internal influence, such as imitation or word of mouth \cite{Bass1969}. Empirical studies of online social-network growth \cite{TrusovBucklinPauwels2009}, viral product features \cite{AralWalker2011}, health-behavior diffusion \cite{Centola2010,Centola2011}, and residential photovoltaic adoption \cite{BollingerGillingham2012} have documented significant peer-influence or word-of-mouth effects. The present model isolates this internal component and adds finite learning and retention memory.

The main limitations follow directly from this baseline design. First, the model is homogeneous: all agents share the same learning and retention thresholds. In the present study, this is a deliberate choice to make the mechanisms transparent, but it also produces abrupt transitions and cohort-like pulses that would likely be smoothed by heterogeneity in technical ability, motivation, opportunity, switching tolerance, or retention. Second, the population is well mixed. Real adoption processes often unfold on structured social, organizational, or geographical networks, where clusters, weak ties, opinion leaders, and local exposure can strongly affect diffusion. 
Previous network extensions of the Bass model provide a natural basis for relaxing the well-mixed assumption \cite{BertottiBrunnerModanese2016BassNetworks,BertottiModanese2019Takeoff,DiLucchioModanese2024}. Incorporating the finite-memory learning, retention, and multi-homing mechanisms introduced here into such networked settings would allow structural and behavioral effects to be studied jointly.
Third, continuing external influence is not modeled explicitly after the initial seed. Advertising, subsidies, prices, procurement mandates, regulation, product redesign, platform governance, and strategic firm behavior may all alter adoption and retention during the diffusion process. Fourth, the model contains only two options and no non-user state, so it is best interpreted as a model of substitution within an established product or behavioral category rather than category creation from non-use.


These limitations point to natural extensions. Fixed agent heterogeneity could be introduced through individual learning thresholds and retention windows. Network structure could replace well-mixed sampling to study organizational segmentation, local lock-in, and bridge users. Ongoing advertising or policy intervention could be represented as an exogenous exposure term, as time-dependent adoption thresholds, or as a direct shift in post-adoption usage preference. Finally, ecosystem dynamics could be modeled more explicitly by allowing key parameters that are fixed in the present baseline to evolve endogenously. Complementary investments could reduce the challenger adoption threshold \(K_Y\), modify the incumbent retention burden \(M_X/T_M\), or alter the compatibility and teaching effects associated with dual adopters.

In summary, the model shows that finite-memory learning and retention at the user level are sufficient to generate aggregate trajectories resembling the four Adner--Kapoor substitution regimes. Its contribution is not to replace ecosystem explanations, but to demonstrate that similar market-level signatures may also emerge from microscopic adoption and retention dynamics. The resulting framework provides a compact way to connect aggregate substitution curves with observable user-level processes: adoption, multi-homing, switching, reversion, and failed learning.

\section*{Data availability}

The data that support the findings of this article are not publicly available upon publication because it is not technically feasible and/or the cost of preparing, depositing, and hosting the data would be prohibitive within the terms of this research project. The data are available from the authors upon
reasonable request.

\bibliography{references}

\end{document}